\documentclass{article}
\usepackage{spconf}
\usepackage{cite}
\usepackage{amsmath,amssymb,amsfonts}
\usepackage{graphicx}
\usepackage{textcomp}
\usepackage{cite}
\usepackage{bbold}
\usepackage{ifpdf}
\usepackage{times}
\usepackage{layout}
\usepackage{float}
\usepackage{afterpage}
\usepackage{amsmath}
\usepackage{amstext}
\usepackage{amssymb,bm}
\usepackage{latexsym}

\usepackage{graphicx}
\usepackage{amsmath}
\usepackage{amsthm}
\usepackage{graphicx}
\usepackage[center]{caption}
\usepackage{pstricks}
\usepackage{caption}
\usepackage{subcaption}
\usepackage{booktabs}
\usepackage{multicol}
\usepackage{lipsum}

\usepackage{enumitem}
\usepackage{amsmath}
\usepackage{algpseudocode}
\usepackage{amsthm}
\usepackage{dsfont}
\usepackage{algorithm}
\usepackage{thmtools}
\usepackage{amssymb}

\def\ba{\boldsymbol{a}}

\def\bg{\boldsymbol{g}}

\def\bu{\boldsymbol{u}}

\def\bx{\boldsymbol{x}}
\def\by{\boldsymbol{y}}
\def\bz{\boldsymbol{z}}

\def\bA{\boldsymbol{A}}

\def\bG{\boldsymbol{G}}

\def\bI{\boldsymbol{I}}

\def\bGamma{\boldsymbol{\Gamma}}

\def\bTheta{\boldsymbol{\Theta}}

\def\bphi{\boldsymbol{\phi}}

\theoremstyle{definition}

\algnewcommand\algorithmicinput{\textbf{Input:}}
\algnewcommand\Input{\item[\algorithmicinput]}
\algnewcommand\algorithmicoutput{\textbf{Output:}}
\algnewcommand\Output{\item[\algorithmicoutput]}
\algnewcommand\algorithmicinit{\textbf{Initialize:}}
\algnewcommand\Init{\item[\algorithmicinit]}

\def\BibTeX{{\rm B\kern-.05em{\sc i\kern-.025em b}\kern-.08em
    T\kern-.1667em\lower.7ex\hbox{E}\kern-.125emX}}
\begin{document}

\title{UPR: A Model-Driven Architecture for Deep Phase Retrieval
}

\name{Naveed~Naimipour $^{\star\mathsection}$, Shahin~Khobahi $^{\mathsection}$, and Mojtaba~Soltanalian \thanks{This work was supported in part by U.S. National Science Foundation Grants CCF-1704401 and ECCS-1809225.}  \thanks{$^{\mathsection}$ The first two authors contributed equally to this work.} \thanks{$^{\star}$Corresponding author (email: \textit{nnaimi2@uic.edu}).}}
\address{Department of Electrical and Computer Engineering,
University of Illinois at Chicago,
Chicago, IL}

\maketitle

\begin{abstract}
The problem of phase retrieval has been intriguing researchers for decades due to its appearance in a wide range of applications. The task of a phase retrieval algorithm is typically to recover a signal from linear phase-less measurements. In this paper, we approach the problem by proposing a hybrid model-based data-driven deep architecture, referred to as the \emph{U}nfolded \emph{P}hase \emph{R}etrieval (\emph{UPR}), that shows potential in improving the performance of the state-of-the-art phase retrieval algorithms. Specifically, the proposed method benefits from versatility and interpretability of well established model-based algorithms, while simultaneously benefiting from the expressive power of deep neural networks. Our numerical results illustrate the effectiveness of such hybrid deep architectures and showcase the untapped potential of data-aided methodologies to enhance the existing phase retrieval algorithms.
 
\end{abstract}


\section{Introduction}
The quest to solve decades old phase retrieval problem has led to numerous algorithms and methodologies. This is no surprise given the many applications of phase retrieval, including those in areas such as crystallography, optics, and imaging \cite{millane,kim}. With an increase in the number of applications in various fields, the developed methodologies continue to increase in number, complexity, and efficiency. Note that a large number of methods in the literature have their roots in the seminal works by Gerchberg, Saxton, and Fienup \cite{GS1,GS2,fienup1,fienup2,fienup3}. However, the Gerchberg-Saxton algorithm's shortcomings in terms of finding the optimal solution in an efficient manner has resulted in numerous new directions. One such direction views the problem from a non-convex lens. In such a scenario, methodologies such as Wirtinger flow (WF), truncated Wirtinger flow (TWF), reshaped Wirtinger flow (RWF), and incremental truncated Wirtinger flow (ITWF) have all shown promise in addressing the problem in an efficient and accurate manner \cite{candes,chen,TWF,RWF}. However, like most established phase retrieval algorithms, they struggle with classical parameter optimization such as determining the optimal step size. On a relevant note, data-driven approaches, such as deep learning techniques, have become immensely useful in recent years for handling complex data sets. Nevertheless, a prevailing issue with these purely data-driven approaches is gaining much more expressive power at the cost of losing interpretability. Although deep learning has been used to perform phase retrieval, it has primarily focused on designing neural nets for limited algorithms such as hybrid-input-output (HIO) and Fienup's method \cite{IntroDNN,Isil}. Such works are limited by their inability to deal with multiple types of models, as well as the more recent phase retrieval algorithms that are more complex in nature. Additional body work has been done using convolutional neural nets, such as prDeep \cite{metzler}, leading to a separate class of architectures not versatile enough to improve the existing algorithms.

The deep unfolding technique is a game-changing fusion of model-based and data-driven approaches. Specifically, it allows for designing model-aware deep architectures based on well-established iterative signal processing techniques. Deep unfolded networks have shown great promise in various signal processing applications \cite{farsad2020data,bertocchi2019deep,khobahi2019deep,shlezinger2020deepsic,khobahi2019model,balatsoukas2019deep,shlezinger2019viterbinet,agarwal2020deep} and are a perfect example of hybrid models that can make use of the immense amounts of data along with utilizing domain knowledge of the underlying problem at hand. Moreover, they can take advantage of the expressive power of deep neural networks, while simultaneously taking advantage of the adaptability and reliability of model-based methods. This makes them an ideal candidate for problems such as phase retrieval, particularly in non-convex settings that struggle with bounding complexity.

In this paper, we propose model-aware deep architectures for the problem of phase retrieval. In particular, we focus on the different variants of the RWF algorithm that have recently shown immense promise in the context of phase retrieval. 
\section{System Model}
\label{sec:model}
The task of phase retrieval is concerned with recovering a complex or real-valued signal of interest, $\bx\in\mathds{R}^{n}/\mathds{C}^{n}$, from $m$ linear phase-less measurements of the form
\begin{eqnarray}
\label{eq:acq}
y_i = |\langle\ba_i, \bx\rangle|,\,\, \mathrm{for}\,\,i  \in \{ 1, 2, \cdots, m\},
\end{eqnarray}
where the set of sensing vectors $\{\ba_i \in \mathds{R}^{n}/\mathds{C}^{n}\}_{i=1}^{m}$, are assumed to be known \emph{a priori}. Define the sensing matrix as
\[
\bA \triangleq  [\ba_1, \ba_2, \cdots, \ba_m]\in\mathds{R}^{n\times m}/\mathds{C}^{n\times m},
\]
with the corresponding measurement vector  given by $\by = [\by_1,\by_2, \cdots, \by_m]^T$. Then, by considering the least-square criterion, the task of recovering the signal of interest, $\bx$, from the measurements vector, $\by$, can be expressed as \cite{RWF}:
\begin{eqnarray}
    \label{eq:opt}
    \underset{\bx\in\mathds{R}^{n}/\mathds{C}^{n}}{\min}f(\bx;\bA,\by)=\frac{1}{2m}\Big\|\by - |\bA^H\bx|\Big\|_2^2.
\end{eqnarray}
Evidently, the problem of phase retrieval is non-convex, and, as described in the previous section, researchers have considered several methodologies to approach this problem. Most notable model-based approaches consider either a loss function of the form \eqref{eq:opt} or an equivalent representation that usually involves higher-order variables. Then, convex methodologies can be utilized to reformulate the problem as a semi-definite program or one can resort to non-convex methods to  tackle \eqref{eq:opt} directly. On the other hand, the existing data-driven approaches make use of the expressive power of deep neural networks and consider a conventional fully connected neural network $f_{\bTheta}(\by)$ (with $\bTheta$ denoting the network parameters), and train it in a manner that the resulting network acts as an estimator of the true signal given the measurements vector $\by= |\bA^H\bx|$. Usually, such data-driven approaches require a large amount of data for training purposes, and more importantly, once trained, it lacks from the inherent interpretability that comes with the model-based approaches. Hence, in this paper we aim to bridge the gap between the model-based and data-driven approaches by proposing a novel model-aware deep architecture based on the well-established first-order optimization algorithms specialized for tackling \eqref{eq:opt}. The resulting network can be seen as a hybrid model-based and data-driven first-order method that enjoys from the interpretability and versatility of model-based algorithms, and at the same time, offers the expressive power of deep neural networks. Moreover, due to the incorporation of the domain knowledge in the deep architecture, it has significantly less trainable parameters and requires much less data to be trained compared to the conventionally `bulky' deep neural networks. To this end, we consider the Incremental Reshaped Wirtinger Flow (mini-batch IRWF) algorithm as a blue-print to design a model-aware deep architecture.

The iterations of the IRWF algorithm for finding the critical points of the non-convex problem in \eqref{eq:opt} can be simply stated as follows: starting from a proper initial point $\bx_0$ (more on this below), the IRWF algorithm generates a sequence of points $\{\bx_0, \bx_1, \bx_2, \cdots\}$ according to the following update rule:
\begin{eqnarray}
    \bx_{k+1} = \bx_{k} - \delta\nabla f(\bx_k;\bA,\by),
\end{eqnarray}
where $\delta$ is some positive step-size, and $\nabla f(\bx_k;\bA,\by)$ denotes the gradient of the objective function in \eqref{eq:opt} realized at the point $\bx_k$, which is given by:
\begin{eqnarray}
    \label{eq:iteartions}
    \nabla f(\bx_k;\bA,\by) = \bA^{H}\left( \bA\bx_k - \by\odot\mathrm{Ph}(\bA\bx_k)\right),
\end{eqnarray}
where the function $\mathrm{Ph}(\bz)$ is applied element-wise and captures the phase of the vector argument; e.g., for real valued signals $\mathrm{Ph}(\bz) = \mathrm{sign}(\bz)$. Due to the non-convex nature of \eqref{eq:opt}, the reconstructed signal can only be recovered up to a global phase difference, and hence, a proper metric to quantify the quality of the reconstructed signal $\bx_{L}$ (by performing $L$ iterations of the form \eqref{eq:iteartions}) can be defined as follows:
\begin{eqnarray}
    \mathrm{D}(\bx_L,\bx^{\star}) \triangleq \underset{\phi\in[0,2\pi)}{\min}\,\|\bx_Le^{-j\phi} -\bx^{\star}\|,
\end{eqnarray}
where $\bx^{\star}$ denotes the true signal. Note that for a real-valued \emph{one-bit phase} scenario, the above metric becomes $\min\|\bx_L ~\pm~ \bx^{\star}\|$. Inspired by \cite{RWF}, we will focus on a real-valued phase retrieval scenario for the rest of this paper. However, the proposed method can be easily extended to complex-valued signals by proper transformations, that will be considered in a future publication.

Generally, the two critical aspects in tackling a non-convex optimization problem (aside from having a proper solver) are: \emph{i}) the choice of the initial starting point for the iterative optimizer, and \emph{ii}) a proper step-size design scheme such that it guarantees the convergence of the sequence $\{\bx_0, \bx_1, \cdots\}$ to a critical point and provides the ability to control and optimize the convergence factor of the underlying iterative solver. Note that the \emph{rate of convergence} of a first-order method cannot be improved unless by resorting to the higher order information. However, the \emph{convergence factor} of such methods can be enhanced and improved by properly tuning the step-sizes resulting in accelerated iterations. A popular choice for obtaining a good initial point for the problem of phase retrieval is known as the spectral method in the literature \cite{candes}. In this paper, we adopt the alternative initialization proposed in \cite{RWF} which benefits from a lower-complexity than that of the spectral method. In particular, the starting point is initialized as $\bx_0 = \lambda_0\bz$, where $\lambda_0\approx\sqrt{\pi/2}$, and $\bz$ is the leading eigenvector of the matrix $\bG = (1/m)\sum_{i=1}^{m} y_i\ba_i\ba_i^H$.

In the following, we present our deep \emph{U}nfolded \emph{P}hase \emph{R}etrieval (\emph{UPR}) framework--- a \emph{model-aware} deep architecture specifically tailored for the problem of phase retrieval, based on the RWF algorithm.

\section{UPR: The Proposed Framework}
In this section, we present the proposed hybrid model-aware and data-driven deep architecture for the problem of phase retrieval. Particularly, we consider the iterations of the form~\eqref{eq:iteartions} as a base-line and unfold them onto the layers of a deep neural network. Before we present the proposed methodology, we first introduce some general concepts from the theory of first-order mathematical optimization. 

Iterative optimization techniques are a popular choice for both convex or non-convex programming. In particular, first order methods are among the most popular and well-established iterative optimization techniques due to their low per-iteration complexity and efficiency in complex scenarios. However, first-order methods generally suffer from a slow speed of convergence and predicting the number of iterations required for convergence is generally a difficult task. As a result, they are not ordinarily suitable for real-time signal processing applications. Consequently, it is natural to consider fixing the total number of iterations of such algorithms along with seeking to optimize the parameters in the iterations that result in the best improvement in the underlying objective function at hand, while allowing only $L$ iterations. Thus, our goal is to improve the existing first order methods by \emph{meta optimizing} the IRWF iterations when the total computational budget is fixed (i.e., allowing only $L$ iterations of the form~\eqref{eq:iteartions}). In order to do so, we formulate the meta-optimization problem in a deep learning setting, and interpret the resulting unrolled iterations as a neural network with $L$ layers where each layer is designed to imitate one iteration of the original iterative optimization method. Eventually, such a deep neural network can be trained using a small data-set and the resulting network can be used as an enhanced first-order method for solving the underlying problem at hand.

Consider the minimization of an objective function $f(\bx)$ and let $\bg_{\phi_i}:\mathds{R}^n\mapsto\mathds{R}^n$ be a parameterized mapping operator defined as
\begin{eqnarray}
\label{eq:mapping}
    \bg_{\phi_i}(\bz) = \left(\mathrm{Id} - \bG_i\nabla f\right)(\bz) = \bz - \bG_i\nabla f(\bz),
\end{eqnarray}
where $\mathrm{Id}$ denotes the identity operator, $\bG_i$ is a positive definite matrix, and $\bphi_i=\{\bG_i\}$ denotes the set of parameters of the operator $\bg_{\bphi_i}$. Then, most of the first-order optimization methods can be represented in terms of the above mapping operator and by considering an updating rule of the form:
\begin{eqnarray}
\label{eq:updaterule}
\bx_{k+1} = \bg_{\phi_k}(\bx_k).
\end{eqnarray}
Specific to our problem is the fact that one can obtain the iterations of the form \eqref{eq:iteartions} by setting $\bG_i=\delta\bI$ for $i\in\{0,1,\dots\}$ and replacing the gradient in \eqref{eq:mapping} with the gradient of the quadratic loss function defined in \eqref{eq:iteartions}, and by following the above updating rule. Generally, for a given set of (pre-conditioning) positive definite matrices $\{\bG_i\}$, the iterations of the form \eqref{eq:updaterule} can be seen as a pre-conditioned gradient-descent method. In this paper, we focus our attention to learning a set of $L$ pre-conditioning matrices $\{\bG_i\}_{i=0}^{L-1}$ where each matrix has a diagonal structure with positive entries. Particularly, we consider $\bG_i=\mathrm{Diag}\left(\delta_i^0, \delta_i^1, \cdots, \delta_i^{n-1}\right)$, where $\delta_i^{j}>0$, $\forall~i,j$. In such a setting, given an initial starting point $\bx_0$, performing $L$ iterations of the form \eqref{eq:updaterule} corresponds to the following composite mapping:
\begin{eqnarray}
    \label{eq:nn}
    \bx_{L} = \mathcal{G}_{\bGamma}(\bx_0) \triangleq \bg_{\phi_{L-1}} \circ \bg_{\phi_{L-2}} \circ \cdots \circ \bg_{\phi_{0}} (\bx_0)
\end{eqnarray}
where $\bGamma = \{\bG_0,\bG_1,\cdots,\bG_{L-1}\}$ represents the overall set of parameters of the mapping operator. From another perspective, the above composite mapping can be seen as a deep neural network with $L$-layers and $\bx_0$ as its input. The output of such a deep architecture is the estimated point after performing $L$ iterations of the underlying iterative optimization algorithm. Thus, the training of such a model-aware deep architecture corresponds to learning the pre-conditioning matrices resulting in an accelerated first-order method. For the training, we first fix the class of the underlying objective function, i.e. by fixing the measurement matrix $\bA$ in our case. Then, we generate a set of observations $\{\by^{(i)} = |\bA^H\bx^{(i)}|\}$ from some known vectors ${\bx^{(i)}}$, and seek to learn the parameters of the network by minimizing the distance between the output of the network $\bx_{L}=\mathcal{G}_{\Gamma}(\bx_0)$ and the optimal points $\bx^{(i)}$ of the fixed class of the objective function. Note that the set of points $\{\bx^{(i)}\}$ are the global minimums of the underlying objective function $f(\bx)$.

\begin{figure*}[!htb]
\centering
\minipage{0.305\textwidth}
   \includegraphics[width=\linewidth]{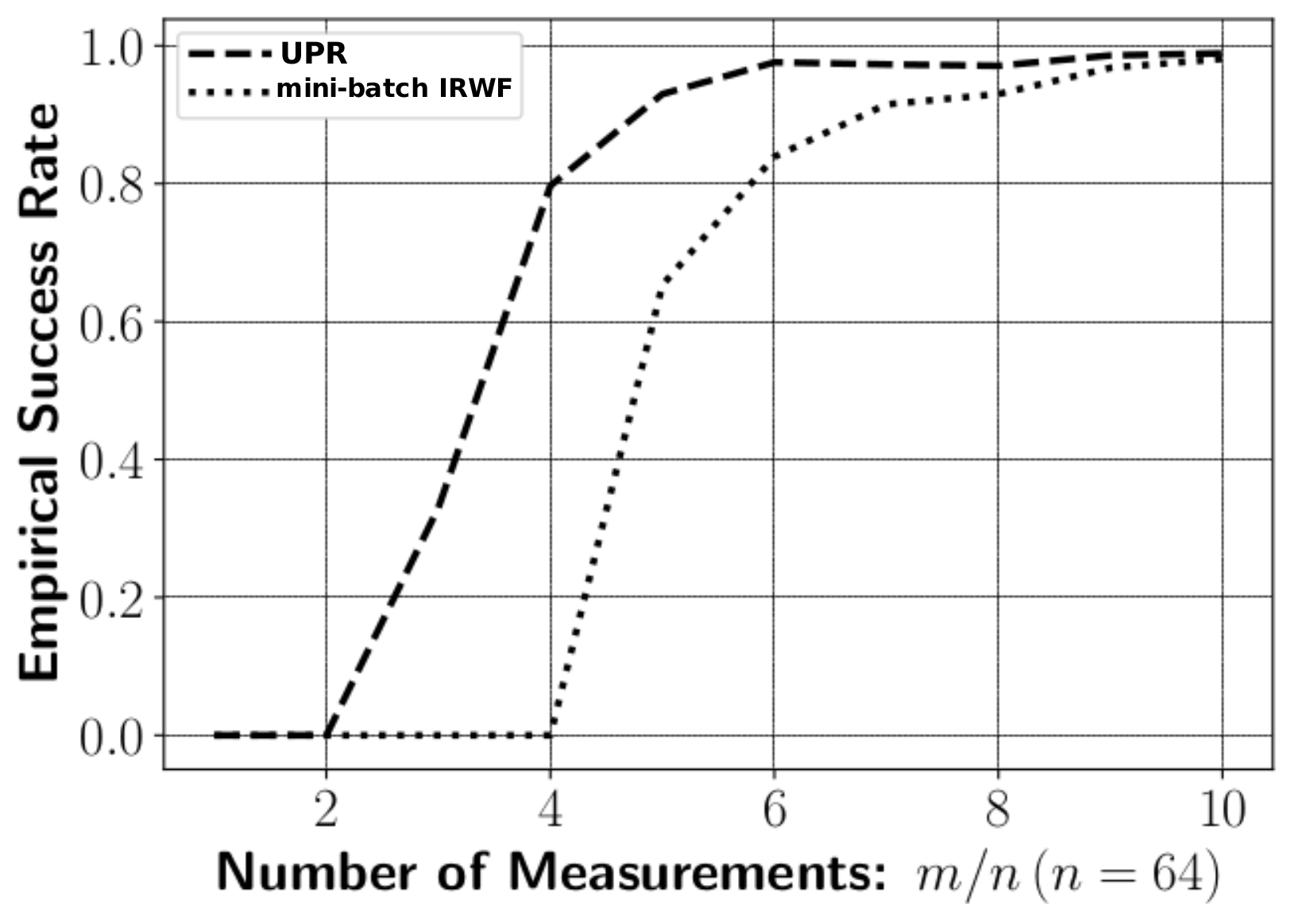}
  \centering
  \subcaption{(a)}\label{fig:a}
\endminipage\hfill
\minipage{0.32\textwidth}
  \includegraphics[width=\linewidth]{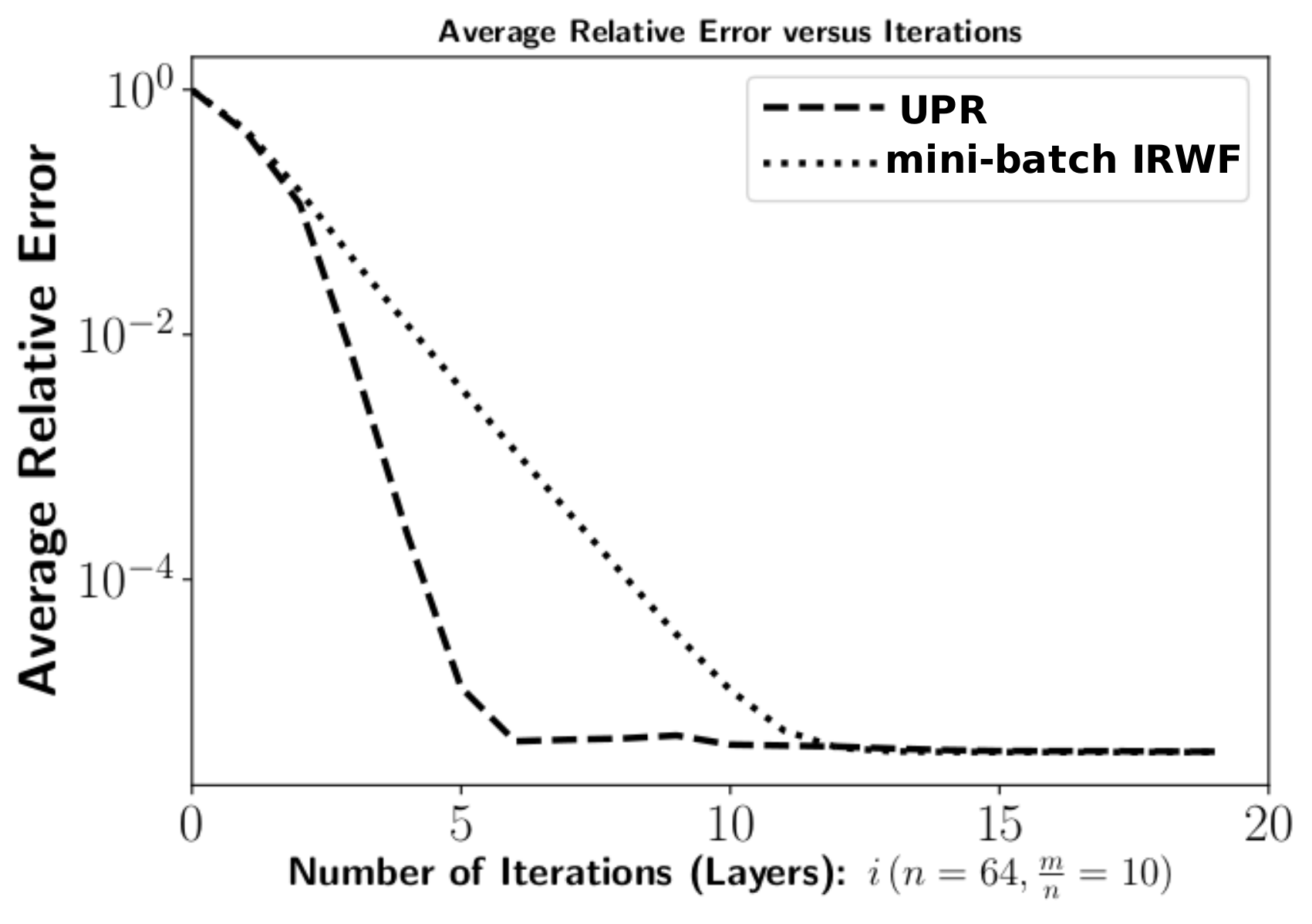}
  \centering
  \subcaption{(b)}\label{fig:b}
\endminipage\hfill
\minipage{0.32\textwidth}%
  \includegraphics[width=\linewidth]{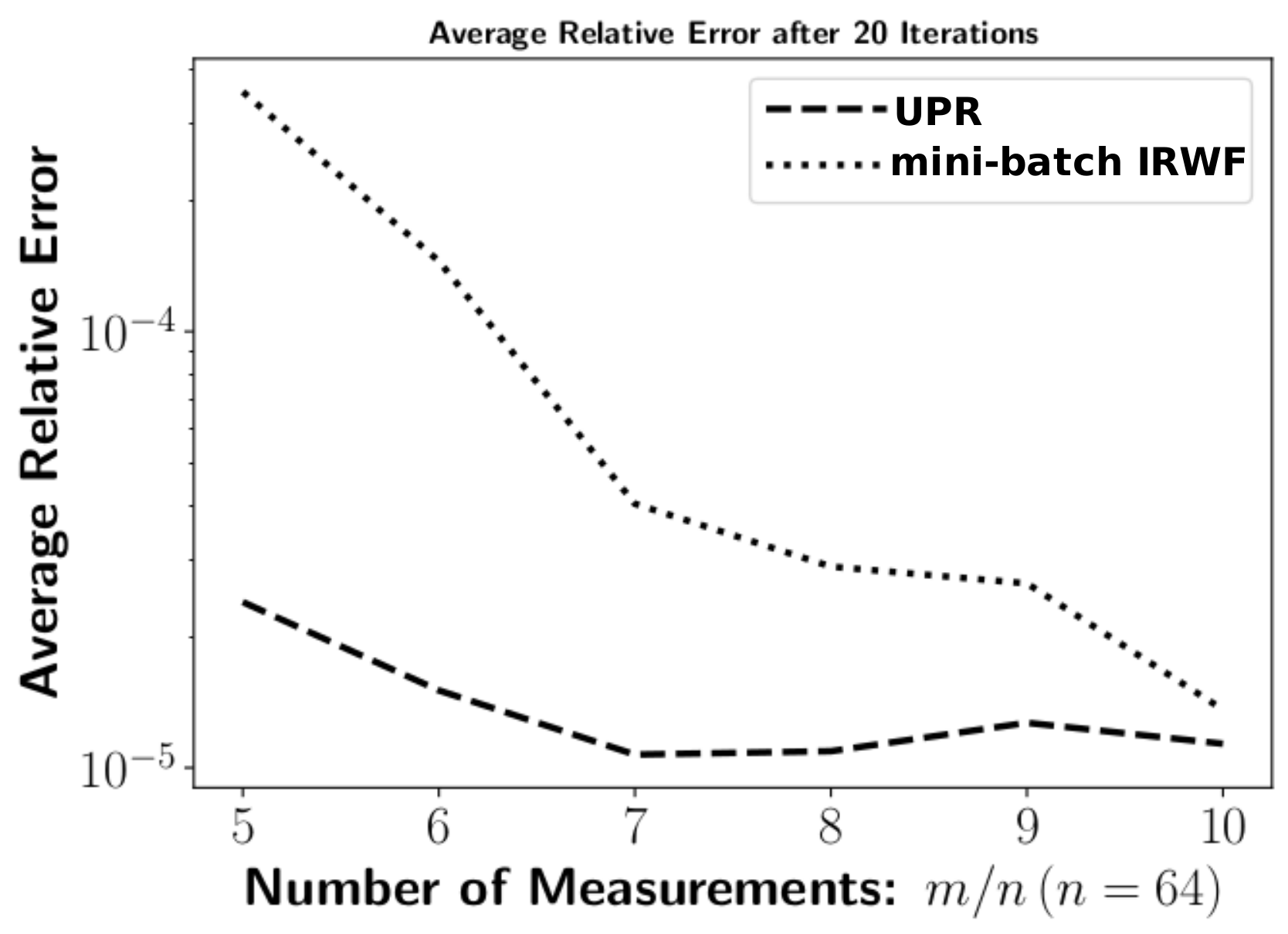}
  \centering
  \subcaption{(c)}\label{fig:c}
\endminipage
\caption{\small Comparison of UPR and mini-batch IRWF: (\textbf{a}) illustrates the empirical rate of success for the two algorithms with respect to $m/n$. (\textbf{b}) shows the average relative error versus the number of iterations (layers) for both algorithms. (\textbf{c}) demonstrates the average relative error versus the number of measurements $m/n$ with $n=64$. It can be observed that the proposed model-aware deep architecture (UPR) significantly outperforms the mini-batch IRWF algorithm in all cases.}
\end{figure*}

In light of the above description, we now present the mathematical structure of the proposed UPR architecture. We define the computational dynamics of the $l$-th layer of the UPR architecture as follows:
\begin{align}
    \bg_{\phi_i}(\bz) &= \bz - \bG_i \bA^{H}\bu, \text{ with}\\
    \bu &= \left(\bA\bz - \by\odot\tilde{\mathrm{sign}}(\bA\bz)\right),
\end{align}
where $\tilde{\mathrm{sign}}(\bx) = \mathrm{tanh}(c\cdot\bx)$, for some large $c>0$, represents an smooth approximation of the $\mathrm{sign}$ function to allow for back-propagating the gradients during the training, $\bz$ is the input to the $l$-th layer, and $\phi_i = \{\bG_i = \mathrm{Diag}(\delta^{0}_{i}, \cdots, \delta^{n-1}_{i})\}$. Hence, the dynamics of the overall network with $L$ layers will be the same as \eqref{eq:nn}.  We consider the training of the proposed architecture via the following optimization problem:
\begin{align}
    \underset{\bGamma}{\min} \left\{\frac{1}{B}\sum_{i=0}^{B}\min\big\|\mathcal{G}_{\bGamma}\left(\bx_0^{(i)}\right)\pm\bx^{(i)} \big\|_2^2\right\},
\end{align}
where $B$ denotes the total number of training points, the training points $\{(\bx^{(i)}, \by^{(i)})\}$ are generated from the data-acquisition model in \eqref{eq:acq}, and the initial points $\bx_0^{(i)}$ are generated using the initialization method described in Section~\ref{sec:model}.

\section{Numerical Results}
In this section, we investigate the performance of the proposed UPR framework for the task of phase retrieval through various numerical experiments. We implemented the proposed UPR architecture using the $\mathrm{PyTorch}$ library\cite{paszke2017automatic}. In addition, for training purposes, we utilized the Adam optimizer \cite{kingma2014adam} with a learning rate of $10^{-3}$. We consider the Empirical Success Rate (ESR) and the average relative error for comparison purposes. Specifically, the ESR metric is defined as the number of successful trials out of $100$ attempts, where a successful trial constitutes obtaining $\mathrm{D}(\bx,\bx^\star)\leq 10^{-4}$, and we define the relative error as $\mathrm{D}(\bx,\bx^\star)/\|\bx^{\star}\|$. In all the simulations, we consider a UPR architecture with $L=20$ layers and also we compare our results with the state-of-the-art mini-batch IRWF algorithm. It should be noted that for a fair comparison, we use the same parameters for the min-batch IRWF method as reported in \cite{RWF}, and consider performing $20$ iterations of it. Both methods are initialized using the spectral method defined in the previous sections. The UPR architecture was trained on a small training data-set of size $B=64$ with $\ba_i\sim\mathcal{N}(\mathbf{0},\bI)$ and $\bx\sim\mathcal{N}(\mathbf{0},\bI)$ for generating both test and training data-sets. The simulations provided in this section are based on evaluating the network on a test data-set that was not shown to the network during the training.

Fig.~1(a) shows the empirical rate of success for the mini-batch IRWF and the UPR architecture with respect to $m/n$. In addition, Fig.~1(b) demonstrates the convergence rate of the proposed method and the mini-batch IRWF algorithm, averaged over $100$ successful trials for $n=64$ and $(m/n)= 10$. Finally, Fig.~1(c) is a plot of the average relative error with respect to $m/n$.

It is evident from Fig.~1(a) that the proposed method outperforms the state-of-the-art mini-batch IRWF in terms of ESR. Specifically, it can be seen that the UPR quickly achieves a very high ESR as compared to its counterpart even for a small number of measurements $m/n$. It was shown in \cite{RWF} that IRWF operates as the best amongst various well-performing algorithms for phase retrieval and the mini-batch IRWF outperforms them all numerically. Thus, the performance of the UPR architecture is presumably superior to those algorithms as well. 

The average relative error versus the number of iterations (layers) for both algorithms is presented in Fig.~1(b). It can be observed that the proposed method significantly outperforms the mini-batch IRWF algorithm in terms of the speed of convergence, and achieves a very low relative error quickly. It was shown in \cite{RWF} that the mini-batch IRWF algorithm converges with a high probability along with a fewer number of iterations as compared to other methods. Hence, our methodology clearly improves upon the performance of the underlying algorithm by learning the  proper pre-conditioning matrices. It should be noted that the ability of the mini-batch IRWF algorithm to converge with a high probability would enable our deep unfolding method to converge even faster. The average relative error versus the number of measurements $m/n$ with $n=64$ is illustrated by Fig.~1(c). Again, it can be seen that the UPR architecture outperforms the mini-batch IRWF in terms of accuracy and achieves a very low relative error even for a small number of measurements.

It should also be mentioned that one of the many pitfalls of non-convex approaches is the difficulty in showing a newly devised algorithm can consistently converge. The purpose of this paper was to show the potential and value of model-based deep learning techniques for the phase retrieval problem. However, although it is well positioned to improve performance, the resulting model-aware deep architecture relies heavily on such convergence guarantees to function properly, which is the case with mini-batch IRWF algorithm. 
\section{Conclusion}
We considered the problem of phase retrieval and proposed a novel hybrid model-based and data-driven deep architecture, the \emph{UPR} framework, showed that it significantly outperforms the state-of-the-art model-based algorithms. The proposed deep architecture enjoys from a relatively small number of trainable parameters compared to other deep learning-based methods and not only benefits from interpretability and versatility of model-based algorithms, but also from expressive power of data-driven methods.
\bibliographystyle{IEEEbib}
\bibliography{refs}

\end{document}